# $C_n$TAB / Polystyrene Sulfonate Mixtures at Air-Water Interfaces: Effects of Alkyl Chain Length on Surface Activity and Charging State

*Felix Schulze-Zachau and Björn Braunschweig*[*]

Institute of Physical Chemistry and Center for Soft Nanoscience, Westfälische Wilhelms-Universität Münster, Corrensstraße 28/30, 48149 Münster, Germany

*Corresponding author: braunschweig@uni-muenster.de

**ABSTRACT**

Binding and phase behavior of oppositely charged polyelectrolytes and surfactants with different chain length were studied in aqueous bulk solutions and at air-water interfaces. In particular, we have investigated the polyanion poly(sodium 4-styrenesulfonate) (NaPSS) and the cationic surfactants dodecyltrimethylammonium bromide ($C_{12}$TAB), tetradecyltrimethylammonium bromide ($C_{14}$TAB) and cetyltrimethylammonium bromide ($C_{16}$TAB). In order to reveal surfactant/polyelectrolyte binding, aggregation and phase separation of the mixtures, we have varied the NaPSS concentration systematically and have kept the surfactants concentration fixed to 1/6 of their respective critical micelle concentrations. Information on the behavior in the bulk solution was gained by electrophoretic mobility and turbidity measurements, while the surface properties were interrogated with surface tension measurements and vibrational sum-frequency generation (SFG). This has enabled us to relate bulk to interfacial properties with respect to the charging state and the surfactants' binding efficiency. We found that the latter two are strongly dependent on the alkyl chain length of the surfactant and that binding is much more efficient as the alkyl chain length of the surfactant increases. This also results in a different phase behavior as shown by turbidity measurements of the bulk solutions. Charge neutral aggregates that are forming in the bulk adsorb to the air-water interface – an effect that is likely caused by the increased hydrophobicity of $C_n$TAB/PSS complexes. This conclusion is corroborated by SFG spectroscopy, where we observe a decrease in the intensity of O-H stretching bands, which is indicative for a decrease in surface charging and the formation of interfaces with negligible net charge. Particularly at mixing ratios that are in the equilibrium two-phase region, we observe weak O-H intensities and thus surface charging.

## 1. INTRODUCTION

The unique physicochemical properties of polyelectrolyte-surfactant (P/S) mixtures both in bulk solutions and at aqueous interfaces makes them interesting for many applications and have attracted considerable interest.[1–4] Applications range from industrial usage such as in mineral processing and oil recovery to cosmetic and pharmaceutical formulations, e. g. for hair care products or drug delivery.[5–12] In most cases, pure polyelectrolytes do not show a significant surface activity, which is why surfactants are often added to influence interface properties of polyelectrolytes in a targeted way. In this respect, mixtures of oppositely charged polyelectrolytes and surfactants are of special interest.[1,8,13–19]

In applications of soft matter materials, polyelectrolytes are used as flocculants and stabilizers due to electrostatic and entropic interactions, which makes them also ideal candidates for stabilization of colloidal solutions and thin liquid films.[20] Therefore, foams, emulsions and thin-liquid films of polyelectrolytes[21,22] and polyelectrolyte/surfactant mixtures[23–33] have been studied before. Varga and Campbell provide a general description for the physical behavior of oppositely charged polyelectrolyte/surfactant mixtures both for polycation/anionic as well as polyanion/cationic surfactants.[3] The general behavior of the latter mixtures is described by slow precipitation of aggregates in the equilibrium two-phase region, which is characterized by a lack of colloidal stability at bulk compositions close to charge neutralization. In in this equilibrium two-phase region, aggregation and sedimentation is occurring and leaves a depleted supernatant after sufficient waiting time. Consequently, this leads also to a depletion of surface-active material and can cause a so-called "cliff edge peak" in the surface tension isotherm after sufficient equilibration time.[34] Differences are observed in the binding efficiency of surfactants to polyelectrolytes as well as in the ageing behavior of the different S/P mixtures. Abraham et al. observed similar effects for mixtures of NaPSS with DTAB and analyzed the effect of ionic strength on the aggregation and changes in the surface tension isotherms.[35] They developed a model for the prediction of the surface tension of such mixtures, which is based on the surface tension of the pure surfactant and additional bulk measurements.[36] The same system was investigated earlier by Noskov et al.[37] who reported that the dynamic surface elasticity is highly dependent on the S/P mixing ratio and their molecular interactions. Noskov et al. also provide an overview of the surface viscoelasticity of S/P mixtures for a range of different systems.[38]

As an experimental method to study polyelectrolytes at fluid interfaces, neutron reflectometry (NR) has been demonstrated as a powerful tool, particularly for investigations of their surface properties. Consequently, NR was often applied to polyelectrolyte/surfactant mixtures at air-water interfaces.[16,39–44] In fact, the effect of the surfactant chain length on the thickness of NaPSS/$C_n$TAB layers adsorbed to the air-water interface has been studied with NR by Taylor et al.[14] However, they varied the surfactant concentration (as opposed to polyelectrolyte concentration in our work) and did not report on the differences in the binding behavior and the surface charge of polyelectrolyte-surfactant complexes at the air-water interface.

In previous work, we have investigated in detail the interactions between poly(sodium 4-styrenesulfonate) (NaPSS) and cetyltrimethylammonium bromide ($C_{16}$TAB).[4] The used polyelectrolyte was shown to have an inherent surface activity if the concentration is high enough.[45] For NaPSS/$C_{16}$TAB mixtures, charge neutral complexes are observed close to equimolar mixing ratios of the two components. At this concentration, we also observed a minimum in surface tension and charge neutral complexes at the interface, which was evidenced by vibrational sum-frequency generation (SFG) spectroscopy. At a threshold concentration that is slightly above this point of zero net charge, we observed a depletion of material from the interface according to surface tension and ellipsometry measurements, similar to what Campbell and co-workers have seen for mixtures of poly(acrylamidomethylpropanesulfonate) sodium salts (PAMPS) with $C_{14}TAB$.[46] However, it is likely that with a change in the surfactant-polyelectrolyte interactions e.g. by variations of the surfactant alkyl chain length, differences in the binding between surfactants and polyelectrolytes can occur. For that reason we present in this work a first detailed study on the effects of surfactant alkyl chain lengths on equilibrium and non-equilibrium properties of $C_nTAB$/ NaPSS mixtures at air-water interfaces as a function of the mixing ratio and alkyl chain length with n = 12, 14 and 16. Using a multi-technique approach, our study provides new information on the physical chemistry of S/P mixtures in the bulk solution and at air-water interfaces.

## 2. MATERIALS AND METHODS

### 2.1 Materials

Poly(sodium 4-styrenesulfonate) with a molecular weight of 70 kDa (PDI < 1.2, batch no. BCBP3081V), dodecyltrimethylammonium bromide ($C_{12}TAB$, ≥ 99%), tetradecyltrimethylammonium bromide ($C_{14}TAB$, ≥ 99%) and cetyltrimethylammonium bromide ($C_{16}TAB$, ≥ 99%) were purchased from Sigma-Aldrich (now Merck). The surfactants were used with and without a previous purification step in order to analyze the influence of possible impurities on the mixtures. The purification was done by threefold recrystallization in acetone with traces of ethanol and was confirmed by surface tension measurements, which are presented in the electronic supplementary information.

### 2.2 Sample preparation

The glassware which was used to prepare and store samples was cleaned with an Alconox detergent solution (Sigma-Aldrich), rinsed with ultrapure water from a Milli-Q Reference A+ purification system (18.2 MΩ·cm, TOC < 5 ppb) and subsequently stored in a bath of 98% p. a. sulfuric acid (Carl Roth, Germany) with the oxidizer Nochromix (Merck) for at least 12 hours. Afterwards, the glassware was rinsed thoroughly with ultrapure water and dried in a stream of 99.999% nitrogen gas (Westfalen Gas, Germany). Stock solutions were prepared by dissolving the necessary amounts in ultrapure water including subsequent sonication for 30 minutes. All S/P mixtures were prepared by simultaneous mixing of equal volumes of surfactant and polyelectrolyte solutions with double their intended final bulk concentration. This procedure was used in order to avoid kinetically trapped aggregates in the equilibrium one-phase region due to local concentration gradients, which are known to influence the results. More information about the latter effects and the role of different mixing protocols is provided in the work by Mezei et al.[19]

### 2.3 Electrophoretic mobility

For the determination of the electrophoretic mobility $u_\zeta$, we have used a Zetasizer Nano ZSP (Malvern, UK). All samples were measured in backscattering geometry with the detector positioned at a scattering angle of 173°. All measurements were performed on freshly mixed samples in triplicates and the results were averaged.

### 2.4 Turbidity

We performed UV/Vis spectroscopy of all samples using a PerkinElmer LAMBDA 650 UV/vis spectrometer. As a measure for the turbidity of the mixtures we evaluated the optical density (*OD*) at a wavelength of 450 nm since there are no specific absorption bands at this wavelength. The turbidity was measured instantly on the freshly mixed samples. After one day, four days and four weeks, the turbidity of the supernatant was measured in order to demonstrate the sedimentation behavior of the mixtures.

**2.5 Surface tension**

Surface tension measurements were performed with a DSA 100 (Krüss, Germany) pendant drop device. Drops were generated with a syringe at the end of a cannula with a diameter of 1.83 mm. We recorded images of the pendent drop at a rate of 1 Hz and evaluated the surface tension from the drop shape using the Young-Laplace equation. All surface tensions were evaluated after 30 minutes. It is important to keep in mind that the presented data do not represent a steady state for all samples. The surface tension has not always reached an equilibrium value in this time scale. However, using the surface tension after 30 min is still a good measure for the surface activity of the samples and is a commonly used approach e. g. to establish relations between interfacial and foam properties.[4,45]

**2.6 SFG spectroscopy**

Sum-frequency generation (SFG) spectroscopy is an inherently interface-sensitive method that is based on the spatial and temporal overlap of two laser beams at the interface of interest. One of the impinging laser beams is a broadband femtosecond infrared (IR) pulse with the frequency $\omega_{IR}$, while the other is a narrowband picosecond visible pulse with the frequency $\omega_{VIS}$. A third beam with the frequency $\omega_{SF} = \omega_{IR} + \omega_{VIS}$ is generated in a second-order nonlinear optical process and detected afterwards. The intensity of the measured sum-frequency signal depends on the intensities of the two incoming laser beams and on the non-resonant and resonant second-order electric susceptibilities $\chi_{NR}^{(2)}$ and $\chi_{R}^{(2)}$, where $\chi_{NR}^{(2)}$ is mainly caused by electronic excitations at the interface and $\chi_{R}^{(2)}$ arises due to resonant excitation of molecular vibrations. Additional contributions from the third-order electric susceptibility $\chi^{(3)}$ are possible if a static electric field is present in the electric double layer, which originates from the adsorption of charged molecules at an interface.[47–50] The magnitude of this contribution is a function of the Debye length $\kappa$, the wave vector mismatch $\Delta k_z$ and the surface potential $\phi_0$.

$$I_{SF} \propto \left| \chi_{NR}^{(2)} + \chi_R^{(2)} + \frac{\kappa}{\kappa + i\Delta k_z} \chi^{(3)} \phi_0 \right|^2$$
$$= \left| \chi_{NR}^{(2)} + \sum_q \frac{A_q e^{i\theta}}{\omega_q - \omega_{IR} + i\Gamma_q} + \frac{\kappa}{\kappa + i\Delta k_z} \chi^{(3)} \phi_0 \right|^2 \qquad (1)$$

The resonant second-order susceptibility depends on the resonance frequency $\omega_q$, the bandwidth $\Gamma_q$ and the amplitude $A_q$ of the vibrational mode $q$ (with the phase $\theta$), which is a function of the number of molecules at the interface and their orientational average. Without a preferred orientation of molecules, as it is the case in the bulk of centrosymmetric materials, $A_q$ equals zero. The symmetry break at interfaces, however, causes nonzero amplitudes and SFG spectroscopy is therefore inherently interface selective for centrosymmetric materials.

SFG spectra were recorded with a home-built spectrometer that is described elsewhere.[4,51,52] Spectra in the range of C-H and O-H vibrations (2800 to 3800 cm$^{-1}$) were measured by tuning the frequency of the IR beam in seven steps. The total acquisition time for the samples was between 4 and 16 minutes. All spectra presented in this study, were recorded with s-polarized sum-frequency, s-polarized visible and p-polarized IR beams (ssp) and were referenced to the nonresonant SFG signal of a polycrystalline gold film that was cleaned in an air plasma prior to the SFG measurement using ppp polarizations.

## 3. RESULTS AND DISCUSSION

### 3.1 Bulk behavior of NaPSS/C$_n$TAB mixtures

In this section, we address the behavior of C$_n$TAB (n = 12, 14 and 16) surfactant mixtures with sodium polystyrene sulfonate (NaPSS) in the bulk of an aqueous solution. Note that each surfactant concentration was kept constant for all measurements and was distinctly below their critical micelle concentration (CMC), which was reported to be at 0.9 mM, 3.5 mM and 15 mM for C$_{16}$TAB, C$_{14}$TAB and C$_{12}$TAB, respectively.[53] In fact, we have chosen for each surfactant a fixed concentration that is ~1/6 of their respective CMC, and varied the NaPSS concentration throughout this study. The surfactant concentrations we used are summarized in Table 1.

*Table 1: Concentrations of apparent free and bound surfactants in the studied C$_n$TAB/NaPSS mixtures. PZC corresponds to the point of zero net charge in the bulk.*

|  | $c_{NaPSS}$ / mM at PZC = $c_{surf, bound}$ | $c_{surf, free}$ / mM | $c_{bound}/c_{total}$ | S/P ratio at PZC |
|---|---|---|---|---|
| 0.1 mM C$_{16}$TAB | 0.08 | 0.02 | 0.8 | 1.3 |
| 0.5 mM C$_{14}$TAB | 0.14 | 0.36 | 0.28 | 3.5 |
| 2.5 mM C$_{12}$TAB | - | - | - | - |

In Figure 1a, we present the electrophoretic mobilities of C$_n$TAB/NaPSS mixtures as a function of NaPSS concentration and the surfactants' chain length. The amount of surfactant bound to the NaPSS polyelectrolyte, which can be qualitative estimated by the change in electrophoretic mobility $u_\zeta$, is strongly dependent on the length of the surfactants' alkyl chain. In fact, the latter is obvious from a careful analysis of the change in $u_\zeta$ as a function of polyelectrolyte concentration in Figure 1a. Here, in particular, an inspection of the surfactant/polyelectrolyte (S/P) mixing ratio at the point of zero net charge (PZC) where the electrophoretic mobility in the bulk vanishes is useful. Figure 1a shows that the PZC is for C$_{16}$TAB at an S/P ratio of 1.3 and for C$_{14}$TAB at an S/P ratio of ~3.5 while with C$_{12}$TAB the binding is not strong enough to form charge neutral complexes (Table 1). Obviously, substantial differences in the binding of surfactants to the PSS polyelectrolyte chain exist and shift as a function of alkyl chain length. In case of $C_{16}$TAB, a much more efficient binding of the surfactant is observed as the PZC is reached at close to equal molar concentration (S/P ratio of 1.3), while in case of $C_{12}$TAB a substantial excess of the surfactant is needed to reach the PZC in the bulk. This conclusion is corroborated by previous works of Svensson et al. as well as Thalberg et al.[54,55] Table 1 provides a qualitative overview of the surfactants' binding efficiency. Note that all polyelectrolyte concentrations refer to the monomer concentration of NaPSS.

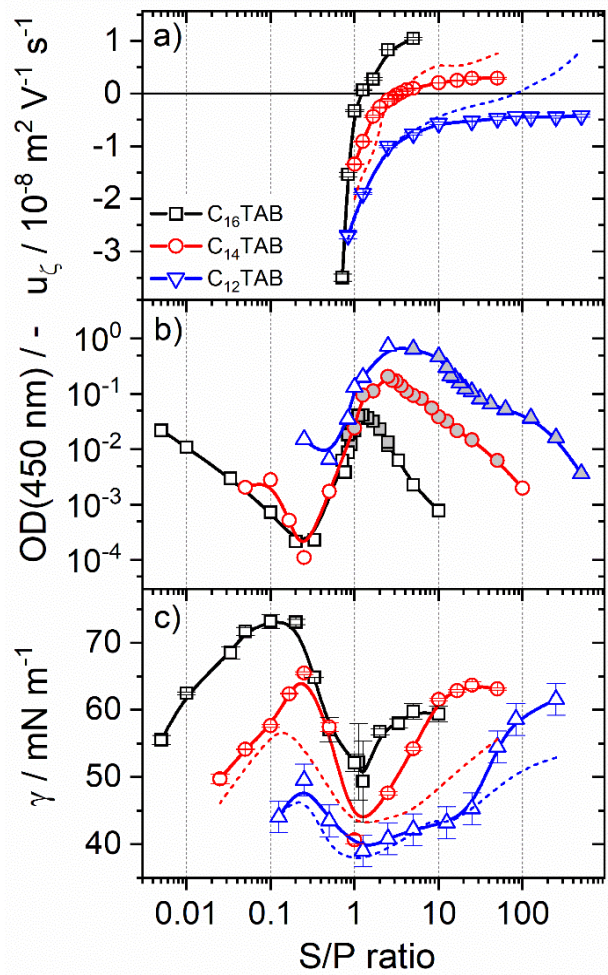

*Figure 1:* Dependence of (a) the electrophoretic mobility $u_\zeta$, (b) the optical density OD at a wavelength of 450 nm directly after mixing (additional results for extended settling times are shown in the ESI) and (c) the interface tension $\gamma$, 30 min after formation of the air-water interface on the $C_nTAB$/NaPSS (S/P) mixing ratio (n= 12, 14, 16 as indicated in (a)). Surfactant concentrations were fixed to 0.1, 0.5 and 2.5 mM for $C_{16}TAB$, $C_{14}TAB$ and $C_{12}TAB$, respectively, while the concentration of NaPSS was varied. Solid lines guide the eye. Dashed lines show the trends of electrophoretic mobility and surface tension of $C_{12}TAB$ and $C_{14}TAB$ without purification. Data points with grey filling mark the equilibrium two-phase region.

Because we can use the S/P ratio at the PZC as a measure of the binding efficiency of $C_n$TAB surfactants to PSS it is interesting to compare the S/P ratio at the PZC for the different $C_n$TAB surfactants. For our further analysis of the binding properties, we make the following assumption: The charge of the polyelectrolyte is compensated by an equal charge of the surfactant, which is accomplished by a 1:1 binding of a surfactant head group to a sulfonate group of the PSS chain. Following the latter approach, we now estimate the apparent amount of bound and free surfactants in solution. For $C_{16}$TAB about 80% of the available surfactant molecules are bound to the PSS chain at the PZC. However, at the PZC of PSS mixtures with $C_{14}$TAB only 30% of the available surfactants bind to PSS, while the remainder contributes to the excess of free surfactants in the solution. For $C_{12}$TAB the PZC cannot be reached with the chosen concentration, meaning that the binding is much weaker compared to to the $C_{14}$TAB. Here, the effect of purification can clearly be seen as the $C_{12}$TAB that was used as received leads to charge neutral and even overcharged complexes. This hints the presence of longer alkyl chains that bind stronger compared to the $C_{12}$TAB. For the unpurified $C_{14}$TAB, we find the PZC at comparable S/P ratio meaning that the general trend is not influenced.

Mixtures of NaPSS and $C_{12}$TAB are well described in previous works: Abraham et al.[35] used 0.48 mM of NaPSS and found charge neutralization at a $C_{12}$TAB concentration of 4.8 mM (S/P = 10), while Varga et al.[3] showed comparable results in the same order of magnitude (6.0 mM $C_{12}$TAB; S/P = 12.5). While there seems to be not a substantial effect of the PSS molecular weight on this binding efficiency,[3] we found that a significantly higher S/P ratio would be necessary to reach the PZC. We have used a different approach (2.5 mM $C_{12}$TAB, variation of the NaPSS concentration) and cannot reach the PZC. These differences with the absolute PSS concentration hint to an additional mechanism for the higher relative surfactant concentration needed for neutralization, when the PSS concentration is low. In fact, we observe that at identical mixing ratios but different total concentrations substantial differences in the binding of the surfactant to the polyelectrolyte exist (Electronic Supplementary Information). This in fact explains the diverse results regarding the binding efficiency. Another possible origin for the latter variations are structural differences, for example, a different degree of sulfonation or a different rigidity of the polyelectrolyte chain. These can arise from different synthesis routes and can have dramatic effects on the polyelectrolyte's physicochemical properties.[56,45] For instance, Abraham et al. discussed that the polyelectrolyte rigidity is related to its the ability to wrap surfactant aggregates and therefore strongly influences the efficiency to bind surfactants.[36] We demonstrate the changes in $u_\zeta$ for different batches of NaPSS and as a function of S/P ratio in the Electronic Supplementary Information.

Figure 1b presents the results of solution's turbidity as a function of S/P mixing ratio and alkyl chain length. For all three systems we observe an equilibrium two-phase region[3] where the samples become turbid due to formation of aggregates that cause light scattering and will eventually sediment on a long-term timescale. This region is indicated in Figure 1b by the grey filled data points, where we define this region when the solution's optical density is decreasing after a sufficient amount of time (e. g. four weeks). The margins of this two-phase region were defined through long-time sedimentation experiments (more information can be found in the ESI).

For mixing ratios that yield a lower mobility, the NaPSS concentration is sufficiently high to provide electrostatic stabilization due to an excess of negative charges. For low NaPSS concentrations or correspondingly for high S/P ratios, overcharged polyelectrolyte/surfactant complexes are stabilized through the excess of positive charges from the surfactants' head groups. In case of $C_{16}$TAB surfactants, the concentration range of the equilibrium two-phase region is within an S/P ratio of around 0.9 and 3.5, while for $C_{14}$TAB it is from S/P ≈ 1.25 to 50 and for $C_{12}$TAB surfactants the equilibrium two-phase region starts at a S/P ratio of 3 and is continuous for higher ratios. In fact, for $C_{12}$TAB surfactants no overcharging is observed, but the turbidity remains on a considerable high level even for S/P = 100. It is important to keep in mind that different surfactant concentrations were used in order to resolve the charge reversal of the polyelectrolyte/surfactant complexes for all three systems; therefore, similar S/P ratios are due to different total NaPSS concentrations for the different systems. This explains why the concentration range of the equilibrium two-phase region increases with surfactant chain length. In the same way, we can explain the increase in optical density of the shorter surfactants.

Comparing the optical density of the mixtures with purified and unpurified surfactants shows that the initially measured turbidity is for all mixtures the same, even for the ones with $C_{12}$TAB where we see significant differences in the electrophoretic mobility. However, this has only consequences for the long-term behavior of these samples. More information is shown in the ESI.

### 3.2 Correlation between bulk and interfacial properties

In order to relate the bulk properties of $C_n$TAB/NaPSS complexes to their surface activity, we have determined the surface tension of S/P mixtures where the NaPSS concentration was varied, while the surfactant concentration was kept constant. Figure 1c presents surface tensions of air-water interfaces that were recorded 30 min after the air-water interface was created.

The changes in surface tension for $C_{16}TAB$/NaPSS mixtures are consistent with a previous report[4] on the latter system and demonstrate that the surface activity is highly dependent on the mixing ratio of NaPSS and $C_{16}TAB$. A close comparison of Figure 1a with Figure 1c clearly shows that at the point of zero net charge (PZC), where negligible electrophoretic mobilities (Figure 1a) in the bulk are found, a distinct minimum in the surface tension is observed (Figure 1c). This observation can be rationalized by a change in hydrophobicity of charge neutral complexes or aggregates that are formed at S/P mixing ratios where the PZC in the bulk is reached. At the mixing ratios where charge neutral complexes are formed, they are more hydrophobic with a higher activity to adsorb at air-water interfaces and thus drive the decrease in surface tension.

$C_{16}TAB$/NaPSS mixtures with negligible mobilities are in the equilibrium two-phase region, where sedimentation at extended waiting times is expected (see above). This is very well documented for different types of oppositely charged S/P mixtures.[3,13,34,57] Further addition of NaPSS and a reduction of the S/P ratio to a value of about 0.1, leads to depletion of $C_{16}TAB$/NaPSS complexes from the air-water interface which is due to their enhanced hydrophilicity at the latter concentrations. These changes with S/P ratio for $C_{16}TAB$/NaPSS mixtures are consistent to what has been reported earlier for $C_{14}TAB$/poly(acrylamidomethylpropanesulfonate) (PAMPS) mixtures.[46]

From a close inspection of Figure 1c, we conclude that the change in surface tension with S/P ratio is for NaPSS mixtures with $C_{14}TAB$ and $C_{12}TAB$ similar to that of $C_{16}TAB$, but consistent with the change in electrophoretic mobility (Figure 1a) the excess of $C_nTAB$ needed to reach the pronounced minimum in surface tension decreases with chain length. Although the changes of the surface tension with S/P ratio are similar for all surfactants, from Figure 1c it is obvious that the surface tensions for surfactants with shorter alkyl chains are systematically lower

($\gamma_{C_{16}TAB} > \gamma_{C_{14}TAB} > \gamma_{C_{12}TAB}$). This can be explained by the higher concentration of both surfactant and polyelectrolyte and the resulting excess of free surfactants that were not bound to the PSS (see Table 1 and corresponding discussion). However, surprisingly, there is a minimum in surface tension for all mixtures at an S/P ratio of 1, which we attribute to the time scale of surface tension measurements. Because the surface tensions presented in Figure 1c were determined 30 minutes after the air-water interface was formed, the results in Figure 1c are strongly influenced by the adsorption kinetics of free surfactants, polyelectrolytes and their complexes. In the equilibrium two-phase region, the adsorption kinetic is very slow, however, with respect to foaming behavior and non-equilibrium properties it is very interesting to see that equimolar mixing leads to fast kinetics. In order to understand the molecular mechanism of the latter, it is helpful to analyze the dynamic surface tension, which provides insight into the adsorption kinetics and is presented in the Electronic Supplementary Information.

Now we want to address the molecular order and charging state of air-water interfaces that were modified by PSS, $C_n$TAB and their complexes by analyzing vibrational SFG spectra (Figure 2).

Vibrational bands which dominate the SFG spectra in Figure 2 and their assignments to vibrational modes are summarized in Table 2.

In the following discussion of the SFG results we want to concentrate first on an analysis of the C-H stretching modes (2800 – 3100 cm$^{-1}$) and later on the O-H stretching modes (3000 – 3800 cm$^{-1}$). Particularly, the ratio of the symmetric CH$_2$ and CH$_3$ stretching vibrations of the pure surfactant solutions is interesting as it provides direct information on the conformation of the surfactants alkyl chains at the interface.[51]

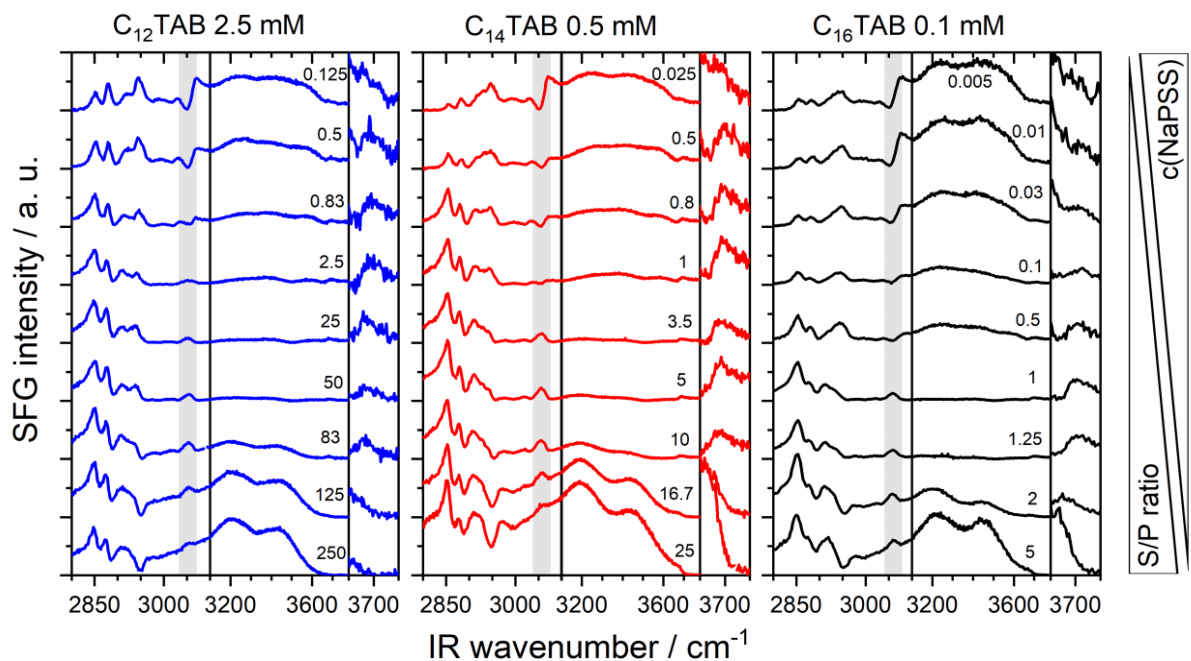

***Figure 2:*** *Vibrational SFG spectra of $C_nTAB$/NaPSS modified air-water interfaces. $C_nTAB$/NaPSS ratios were varied as indicated next to the spectra by varying the NaPSS concentration (0 to 20 mM), while the surfactant concentrations were kept fixed to 2.5, 0.5 and 0.1 mM for $C_{12}TAB$, $C_{14}TAB$ and $C_{16}TAB$, respectively. The spectra have an offset for better visualization. Shown is the region of C-H vibrations (aromatic C-H stretch is highlighted), the region of O-H vibrations and a close-up (10 x) on the dangling O-H vibration.*

While for $C_{16}TAB$ and $C_{14}TAB$ surfactants, no clear difference in the intensity ratio of symmetric $CH_2$ and $CH_3$ stretching vibrations at 2850 and 2878 cm$^{-1}$ is observed, this ratio is significantly different in the SFG spectra from $C_{12}TAB$ solutions at the air-water interface. In fact, for the air-water interfaces from the pure surfactant solutions, a small $CH_2/CH_3$ SFG intensity ratio (< 0.1) of the symmetric stretching bands is indicative for well-ordered alkyl chains at the interface that show a low density of gauche conformations and predominant all-trans conformation. In addition, previous studies[58–60] have shown that alkyl chain ordering is not only a function of surface coverage[59,58] but also a function of the surfactants alkyl chain length.[60] That is because shorter chains lead to a decrease in lateral dispersive interactions and result in lower molecular order and higher density of gauche conformations. In our spectra, however, the highest molecular order was observed for $C_{12}$TAB surfactants. This result can be rationalized by the higher bulk concentration and consequently higher surface excess and a more closed-packed layer. This smaller $CH_2/CH_3$ ratio for $C_{12}$TAB is observed also in the mixtures for S/P ratios > 25, while for low S/P ratios the higher NaPSS concentration leads to similar $CH_2/CH_3$ intensity ratios for all three surfactants. Because polystyrene sulfonate also gives rise to methylene stretching bands, C-H contributions from PSS chains to the SFG spectra need to be considered in the interpretation of the SFG spectra at low S/P ratios, where complexes become the dominant species at the interface, as we will show below. In fact, this explains the close resemblance of the SFG pectra for the mixtures low S/P ratios (Figure 2).

While the C-H vibrations originate from both NaPSS and $C_n$TAB moieties at the air-water interface, the aromatic C-H stretch is a direct signature for the presence of PSS molecules at the air-water interface. Already small additions of 0.05 mM NaPSS (S/P ratio ~5 in case of $C_{16}$TAB) lead to a noticeable vibrational band at 3060 cm$^{-1}$ (Figure 2). In the absence of $C_n$TAB surfactants the SFG spectra of solutions with a low concentration of NaPSS < 2mM were identical to the SFG spectra from neat air-water interface without adsorbed species.[45] While the neat polyelectrolyte is not surface active at low concentrations, the increased surface activity PSS can be explained by a cooperative effect of $C_n$TAB surfactants and PSS polyelectrolytes and is direct evidence for the formation and adsorption of $C_n$TAB/PSS complexes to the air-water interface. These complexes show obviously a higher surface activity than the neat PSS molecules, which comes up to expectations and can be linked to the increased hydrophobicity of PSS molecules in the presence of $C_n$TAB surfactants.

*Table 2: Assignments of vibrational bands.*

| Wavenumber / cm$^{-1}$ | Assigned molecular vibration | Ref. |
|---|---|---|
| 2850 | Methylene symmetric stretch | 52,61,62,59 |
| 2878 | Methyl symmetric stretch | 52,61,62,59 |
| 2920 | Methylene asymmetric stretch | 52,61,62,59 |
| 2946 | Methyl Fermi resonance | 52,61,62,59 |
| 2970 | Methyl asymmetric stretch | 52,61,62,59 |
| 3060 | Aromatic C-H stretch | 52,61,62,59 |
| 3200, 3450 | O-H stretch of interfacial water | 52,63 |
| 3450 | O-H stretch of interfacial water | 52,63 |
| 3710 | "Dangling" O-H | 52,63 |

Interestingly, S/P complexes were observed at the air-water interface for all samples when the NaPSS concentration was adjusted to > 0.05 mM. From the poor surface activity of PSS (see above) and clear signatures of PSS (band at 3060 cm$^{-1}$) in the SFG spectra in Figure 2, we conclude that for low S/P mixing ratios, the S/P complexes are always present at the air-water interface. This conclusion is consistent with the report by Taylor et al.[14] who have proposed "primary complexes" that consist of a surfactant monolayer with the polyelectrolyte backbone attached underneath. While these primary complexes are present at the surface under all circumstances, another kind of secondary complex may attach to the underside of the initial monolayer complex, which is required for stabilization of the surface and a decrease of surface tension. This description of different types of complexes is consistent with our findings especially for the NaPSS/C$_{16}$TAB system. Here we detect complexes at the interface with SFG at an S/P ratio of 0.1, which means that they are present at the interface while they do not lower the surface tension compared to a neat water surface (Figure 2c vs Figure 1c). This matches Taylor's description of "primary complexes" which do not necessarily lower the surface tension. For the two shorter surfactants, at least the trend is similar in a way that there is a local maximum at comparable S/P ratios. Higher S/P ratios lead to a lowering of the surface tension, which is consistent with the observation of secondary complexes as described by Taylor et al. who reported that the latter tend to be more hydrophobic and adsorb at the interface with larger surface excess. Most likely, however, this decreased surface tension around S/P ratios of 1 can be linked to the formation of aggregates. According to the results of our study, these mixing ratios lead to aggregates in the bulk (Figure 1b) – however, the physical behavior in our case is naturally different since we vary the polyelectrolyte concentration and use much higher concentrations in the mixtures. Here, aggregates that were previously formed in the bulk (Figure 1b) are likely to be adsorbed at the interface because of their much higher concentration. At the highest NaPSS concentrations (lowest S/P ratio), a comparison between SFG spectra of the different $C_n$TAB/NaPSS mixtures reveals great similarities in terms of spectral shape and band positions. That is because at low S/P ratios there is an excess of negative charges in the mixtures (Figure 1a). However, the absolute charge density at the air-water interface due to the presence of $C_{16}$TAB/PSS complexes varies between surfactants. This conclusion can be made from a close analysis of the O-H stretching bands in the SFG spectra of Figure 2. We will now explain how the conclusions about the surface charging can be made from the SFG intensity of the O-H stretching bands: Equation 1 indicates that the intensity of the SFG signal is strongly dependent on the surface potential and thus related to the surface charge via the Grahame equation. The probing depth, and with this the number of molecules in the near-interfacial region contributing to the SFG signal, is also a function of the Debye length and therefore the ionic strength. Since the latter two were not constant in our samples, the SFG intensity of O-H stretching modes is partly affected by interference effects and not exclusively related to the surface potential according to equation (1).[50] However, this is only relevant for ionic strengths << 1 mM, and thus care must be taken at high S/P ratios for $C_{16}$TAB and $C_{14}$TAB surfactant. In fact, for most of the concentrations we show such effects are irrelevant and the results reflect qualitative changes of the interfacial net charge. In order to analyze the changes in the interfacial net change in more detail, we have determined the O-H intensities of the low and high frequency branches at 3200 and 3450 cm$^{-1}$ in the SFG spectra of Figure 2 and present the results of this analysis in Figure 3.

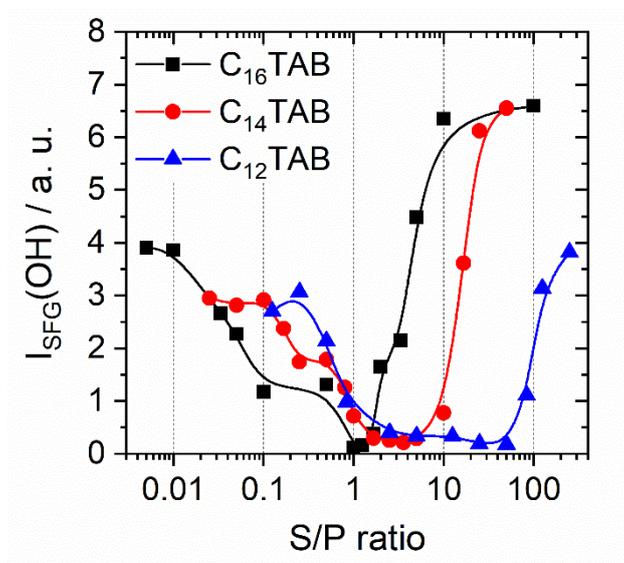

*Figure 3: Intensity of O-H vibrational bands of the measured SFG spectra as a function of the surfactant/polyelectrolyte ratio. Surfactant concentrations are 0.1 mM for C$_{16}$TAB, 0.5 mM for C$_{14}$TAB and 2.5 mM for C$_{12}$TAB. Solid lines guide the eye.*

From a close inspection of Figure 3 and a comparison with Figure 1a, it becomes obvious that the O-H intensity shows a pronounced minimum at S/P mixing ratios, where also the electrophoretic in the bulk vanishes at the bulk PZC. The S/P ratio at the PZC in the bulk as shown in Table 1 is 1.3 for C$_{16}$TAB and 3.5 for C$_{14}$TAB while there is no PZC for C$_{12}$TAB. Mixtures with these ratios also show negligible charging at the interface, which can be deduced from negligible SFG intensities from O-H stretching bands of hydrogen-bonded interfacial water molecules. Obviously, bulk and interfacial properties in terms of electrostatics are for all C$_n$TAB/NaPSS mixtures closely related. Furthermore, from the results in Figures 2 and 3 we can conclude that for mixing ratios close to the PZC, the charge neutral complexes which form aggregates in the bulk (see discussion above) dominate interfacial properties. However, we also find positively charged complexes at the interface for C$_{12}$TAB/PSS mixtures that show negatively charged complexes in the bulk. Compared to the behavior of the mixtures with the longer surfactants, this finding provides new insight and hints two different types of complexes in the mixtures, one of which is negatively charged in the bulk and one of which is positively charged with excess of surfactant and can be found at the interface.

## 4. CONCLUSIONS

Mixtures of oppositely charged polyelectrolyte (NaPSS) and $C_n$TAB surfactants with different alkyl chain lengths (n = 12, 14 and 16) have been studied in the bulk solution and at air-water interfaces. Using vibrational sum-frequency generation to study the composition and charging state of air-water interfaces as well as measurements of the electrophoretic mobility and the solution's turbidity, we have revealed the effects of the surfactant's chain length on their binding behavior to polystyrene sulfonate. Depending on the surfactant/polyelectrolyte (S/P) mixing ratios, we observe aggregation and phase separation in the bulk which is similar to all surfactants but at concentrations which are shifted to higher S/P ratios as length of the alkyl chain decreases. In fact, the point of zero net charge is at S/P mixing ratios of 1.3 and 3.5 for $C_{16}$TAB and $C_{14}$TAB surfactants, while for $C_{12}$TAB the PZC is not crossed for the investigated concentrations. All solutions with S/P mixing ratios which showed electrophoretic mobilities within a range of -1.5 to 0.8 x $10^{-8}$ $m^2$ / (V s) form complexes that are sufficiently hydrophobic to favor aggregation, which leads to turbid solutions and a long-term sedimentation of aggregates (equilibrium two-phase region). Furthermore, we can relate the phase behavior of the mixtures to the hydrophobicity and the resulting increased surface activity of S/P complexes. The margin of the equilibrium two-phase region marks a point of drastic changes within the bulk and interfacial behavior: at lower S/P ratios, complexes still carry sufficient negative charges in order to stabilize the complexes against aggregation, which leads to an increase of surface tension when the S/P ratio is decreased from this point on. Higher S/P ratios however yield poorly charged complexes and aggregates that can enter the surface and lower the surface tension. Further increasing the S/P ratio leads to overcharged complexes that are then again more hydrophilic and increase the surface tension. These effects lead to a minimum of the surface tension at S/P ratios of 1 for all surfactants after 30 min. This can in part be explained by the margin of the equilibrium two-phase region, but also by the non-equilibrium state of the surface tension. In fact, the adsorption kinetics of mixtures around the PZC still show a large ongoing decrease of the surface tension, which leads to lower surface tensions when measured after a longer time, whereas for other mixing ratios longer measurement times will not change the surface tension considerably. SFG spectroscopy shows there are complexes at the interface for all investigated mixtures, yet they do not lead to lowered surface tension in all cases. The charge of the complexes that are present at the air-water interface, as qualitatively estimated using the intensity of O-H stretching vibrations, shows excellent agreement with bulk measurements, which leads to the conclusion that aggregates formed in the bulk can enter the surface due to their hydrophobicity.

## ACKNOWLEDGEMENTS

The authors gratefully acknowledge funding from the European Research Council (ERC) under the European Union's Horizon 2020 research and innovation program (Grant Agreement 638278).

**Graphical abstract**

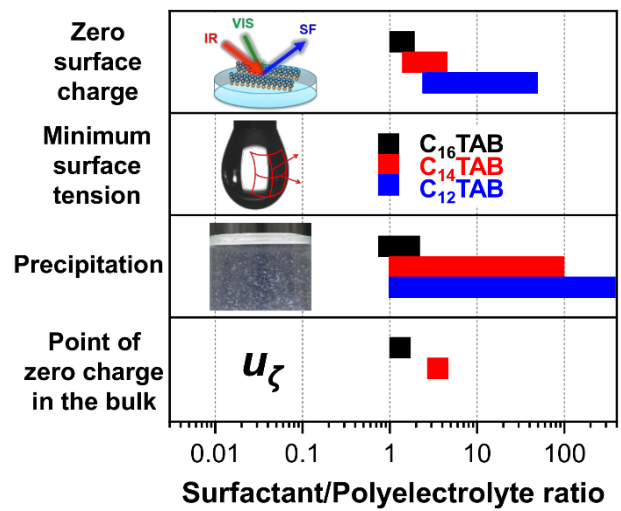

The physical behavior of surfactant/polyelectrolyte mixtures in bulk and at interfaces is displayed in dependency of the surfactant chain length.